\documentclass[12pt]{article}
\usepackage{amsmath,amssymb,exscale,cancel}
\usepackage{slashed}
\usepackage{graphicx}
\usepackage{epsfig}
\usepackage{multicol}
\usepackage{color}
\usepackage{mathrsfs}
\usepackage{blindtext}
 \usepackage{fancyhdr}
\usepackage{hyperref}
\usepackage{cite}
\usepackage{mathtools}

\usepackage{floatrow}

\usepackage{graphicx}
\usepackage{sidecap}

\usepackage[latin1]{inputenc} 

\usepackage{multicol}  
 
 \usepackage{titlesec}
 
 \usepackage{rotating,slashed,xcolor,amsfonts,expdlist,charter}

\numberwithin{equation}{section}

\usepackage{xcolor}
\usepackage{sectsty}

\usepackage{mdframed}
\usepackage{titletoc}

\numberwithin{equation}{section}

\usepackage{xcolor}
\usepackage{sectsty}

\usepackage{mdframed}
\usepackage{titletoc}

\definecolor{secnum}{RGB}{13,151,225}
\definecolor{ptcbackground}{RGB}{212,237,252}
\definecolor{ptctitle}{RGB}{0,177,235}

\titlecontents{lsection}
  [5.8em]{\sffamily}
  {\color{secnum}\contentslabel{2.3em}\normalcolor}{}
  {\titlerule*[1000pc]{.}\contentspage\\\hspace*{-5.8em}\vspace*{5pt}%
    \color{white}\rule{\dimexpr\textwidth-15.5pt\relax}{1pt}}

\usepackage{hyperref}
\hypersetup{colorlinks,bookmarksopen,bookmarksnumbered,citecolor=rossos,
linkcolor=redy,pdfstartview=FitH,urlcolor=green-go}
\usepackage{slashed}

\definecolor{blus}{cmyk}{1,0.9,0,0.1}
\definecolor{verdes}{cmyk}{0.99,0,0.59,0.65}
\definecolor{rossos}{cmyk}{0,1,1,0.55}
\definecolor{redy}{cmyk}{0,1,1,0.7}
\definecolor{greeny}{cmyk}{0.99,0,0.59,0.98}
\definecolor{green-go}{cmyk}{0.79,0,0.59,0.5}

\usepackage{titlesec}

\newcommand{\beq}{\begin{equation}}
\newcommand{\eeq}{\end{equation}}

\def\hhref#1{\href{http://arxiv.org/abs/#1}{arXiv:#1}} 

 \def\Lag{\mathscr{L}}
 
\newcommand{\tmtextbf}[1]{{\bfseries{#1}}}
\newcommand{\tmtextrm}[1]{{\rmfamily{#1}}}
\newcommand{\bp}{\bar M_P}

\def\be{\begin{equation}}
\def\ee{\end{equation}}
\def\ba{\begin{array} }

\def\bac{\begin{array} {c}}
\def\bacc{\begin{array} {cc}}
\def\baccc{\begin{array} {ccc}}
\def\bacccc{\begin{array} {cccc}}
\def\ea{\end{array}}
\def\bea{\begin{eqnarray}}
\def\eea{\end{eqnarray}}

\definecolor{red}{rgb}{1,0,0}

\def\psl{\hbox{\hbox{${p}$}}\kern-1.9mm{\hbox{${/}$}}}
\def\dsl{\hbox{\hbox{${\partial}$}}\kern-2.2mm{\hbox{${/}$}}}
\def\Dsl{\hbox{\hbox{${D}$}}\kern-2.6mm{\hbox{${/}$}}}

\newcommand{\gappeq}{{\rlap{{\raise}.5ex\text{\ensuremath{>}}}{{\lower}.5ex\text{\ensuremath{\sim}}}}}
\newcommand{\lappeq}{{\rlap{{\raise}.5ex\text{\ensuremath{<}}}{{\lower}.5ex\text{\ensuremath{\sim}}}}}
\newcommand{\I}{\tmtextrm{1{\kern}-.24em l}}

\begin{document}
\topmargin -1.0cm
\oddsidemargin 0.9cm
\evensidemargin -0.5cm

{\vspace{-1cm}}
\begin{center}

\vspace{-1cm}

 {\LARGE \tmtextbf{ 
\color{rossos} \hspace{-0.9cm}    \\
Reheating after the Supercooled Phase Transitions 
 \\  with Radiative Symmetry Breaking
 \hspace{-1.6cm}}} {\vspace{.5cm}}\\



\vspace{1.3cm}

{\large{\bf   Francesco Rescigno {\rm and} Alberto Salvio}}

{\em  

\vspace{0.6cm}

Physics Department, University of Rome and INFN Tor Vergata, Italy\\

%
%
%
%
%
%
%
 \vspace{1.5cm}

%
%
%
%
%
}
%
%
\end{center}

%
%
\noindent ---------------------------------------------------------------------------------------------------------------------------------
\begin{center}
{\bf \large Abstract}
\end{center}
\noindent  Theories with radiative symmetry breaking (RSB) lead to first-order phase transitions and the production of gravitational waves as well as primordial black holes if the supercooling period lasted long enough. Here we explain how to efficiently reheat the universe after such period in the above-mentioned class of theories. Two cases are possible, depending on whether the RSB scale is much larger than the electroweak (EW) symmetry breaking scale or not. When it is, the dominant reheating mechanism can be the decays of the field responsible for RSB in the Standard Model (SM) sector. We point out that in a similar way dark matter (DM) can be produced and we analyze in some detail the case of a sterile-neutrino, finding that the full DM abundance is reproduced when this particle is at the $10^2$ MeV scale in a well-motivated SM completion. When the RSB scale is not much larger than the EW symmetry breaking scale, we find that efficient reheating always  occurs when the energy density of the false vacuum is first entirely transferred to a dark photon and then to SM fermions via dark-photon decays.

\vspace{0.7cm}

\noindent---------------------------------------------------------------------------------------------------------------------------------

  \vspace{-0.9cm}
  
  \newpage 
\tableofcontents

\noindent

\vspace{0.5cm}

\noindent ---------------------------------------------------------------------------------------------------------------------------------

\section{Introduction}\label{Introduction}

 Gravitational wave astronomy has become an extremely active and exciting field of physics after the discovery of the gravitational waves from  binary black hole and neutron star mergers~\cite{Abbott:2016blz,TheLIGOScientific:2016wyq1,LIGOScientific:2017ync}. Few years ago the interest in this field was further boosted by the detection of a background of GWs by pulsar timing arrays: these include the North American Nanohertz Observatory for Gravitational Waves (NANOGrav), the Chinese Pulsar Timing Array (CPTA), the European Pulsar Timing Array (EPTA) and the Parkes Pulsar Timing Array (PPTA)~\cite{NANOGrav:2023gor,Antoniadis:2023ott,Reardon:2023gzh,Xu:2023wog}. In turn this has revived the interest in primordial black holes as (part of) the observed DM abundance.
 
Quite generically,  first order phase transitions (PTs) can lead to observable gravitational waves and primordial black holes (see e.g.~\cite{Hawking:1982ga,Crawford:1982yz,Kodama:1982sf,Moss:1994iq,Khlopov:1999ys,Maggiore:2018sht,Kawana:2021tde,Liu:2021svg,Baker:2021sno,Kawana:2022olo,Athron:2023xlk,Lewicki:2023ioy,Gouttenoire:2023naa,Ahriche:2023jdq,Hashino:2025fse,Kierkla:2025vwp}). Such PTs always  occur~\cite{Witten:1980ez,Salvio:2023qgb,Salvio:2023ext,Banerjee:2024cwv,Salvio:2024upo} when symmetries are broken and masses are generated mostly through radiative effects~\cite{Coleman:1973jx,Gildener:1976ih}. 
This RSB scenario also offers the possibility of explaining large hierarchies between mass scales as these are generated through dimensional transmutation~\cite{Gildener:1976ih}. Furthermore, in this type of PTs there is always a long period of supercooling~\cite{Witten:1980ez,Salvio:2023qgb}, when the field $\chi$ responsible for the PT is trapped in the false vacuum before the PT actually takes place. 

If supercooling is strong enough a model-independent description of the PTs and the consequent production of gravitational waves and primordial black holes becomes possible~\cite{Salvio:2023qgb,Salvio:2023ext,Banerjee:2024cwv}. After this very long supercooling period  preexisting matter and radiation are diluted down to a non-observable amount. Therefore, in this situation, bubbles of the true vacuum expand at approximately the speed of light in the false vacuum background.\footnote{For studies of the effect of a preexisting plasma in first-order phase transitions see e.g.~Refs.\cite{Jinno:2019jhi,Gouttenoire:2021kjv}.} However, after supercooling, the universe must be reheated. 

Here we study if and how the universe is reheated  in a general RSB theory, focusing on mechanisms that do not require specific model building, but can be realized in the general RSB scenario. 

 We consider first the reheating mechanism where the particle responsible for RSB directly decays into SM particles and/or particles with sizable couplings to the SM. This mechanism is analogous  to the standard reheating occurring after the original inflationary period, where it is instead the inflaton  that decays into the SM sector (see~\cite{WeinCosmo} for an introduction). However, unlike the original inflationary period, in the case of supercooled PTs the universe must be literally reheated as the PTs are preceded by a hot period, when a non-vanishing temperature was present. This reheating mechanism is particularly effective when the SM is embedded in the RSB theory, meaning that the SM has sizable couplings with $\chi$. In this case the RSB  scale must be significantly larger than the EW scale to satisfy the experimental bounds.
 
It is interesting to note that, besides creating SM particles, reheating can also produce DM particles, which can contribute to the full DM abundance in addition to primordial black holes (PBHs). We will illustrate this point by considering the production of sterile-neutrino DM\footnote{See~\cite{DelleRose:2019pgi,vonHarling:2019gme,Ghoshal:2020vud} for RSB models where DM is instead due to the QCD axion.}.

When the RSB scale is not larger than the EW scale the RSB theory has to be a dark (somewhat hidden) sector. In that case we will explore the possibility of reheating the universe through preheating~\cite{Dolgov:1989us,Traschen:1990sw,Kofman:1994rk,Kofman:1997yn}, when production of particles interacting with $\chi$ occurs as a result of the time dependence of this field through parametric resonance.  This alternative mechanism can indeed produce particles heavier than the field responsible for RSB.  

Other mechanisms explored in specific models can further contribute to particle production (see e.g.~\cite{Watkins:1991zt,Baldes:2023cih,Mansour:2023fwj,Shakya:2023kjf,Giudice:2024tcp,Baldes:2024wuz} for extended analysis of particle production through bubble collisions), increasing the reheating temperature, but in this paper, as already mentioned, we focus on those that can be studied in the general RSB scenario.  

The paper is organized as follows. In Sec.~\ref{General structure of the theory} we discuss the general structure of the RSB scenario. In Sec.~\ref{Interactions} we identify and study the leading interactions that are responsible for the decays of $\chi$ in the SM sector, which, as already mentioned, play a crucial role in reheating the universe when the SM is part of the RSB theory. Sec.~\ref{Decay Rates} (and an appendix) is then devoted to the calculation of all corresponding decay rates.   In Sec.~\ref{sec:RSBEW} we illustrate in a simple, yet prototypical, setup how the EW symmetry breaking (EWSB) is triggered by RSB when the SM is part of the RSB theory. The results of Sec.~\ref{sec:RSBEW}, appropriately generalized, are then used in Sec.~\ref{Reheating} to compute the reheating temperature and to determine sufficient conditions for fast reheating, when the entire energy density stored in $\chi$ is transferred to the SM plasma (other considerations regarding reheating after a first-order phase transition were provided in Ref.~\cite{Gouttenoire:2023pxh}).  Sec.~\ref{Reheating} also investigates when those conditions are compatible with the requirement that the bubbles of true vacuum are not diluted by the rapid expansion of the universe in Sec.~\ref{Compatibility between reheating and phase transition}. Moreover, Sec.~\ref{sec:Neutrini} explains how to produce sterile-neutrino DM through the decays of $\chi$. In order to illustrate how these general results can be used in specific models, we investigate in some detail an RSB SM extension featuring a gauged $B-L$ symmetry and three right-handed (sterile) neutrinos in Sec.~\ref{Radiative electroweak and lepton symmetry breaking}.  Finally, Sec.~\ref{Preheating} provides sufficient conditions to efficiently reheat the universe when the RSB scale is below the EWSB scale, when the RSB theory must be a dark sector. Our conclusions are offered in Sec.~\ref{Conclusions}.

\section{General structure of the theory}\label{General structure of the theory}

We consider the most general no-scale matter\footnote{It is also possible to consider a no-scale gravitational Lagrangian and generate the Planck mass and the cosmological constant radiatively~\cite{Salvio:2014soa,Kannike:2015apa,Salvio:2017qkx,Salvio:2017xul,Salvio:2019wcp,Salvio:2020axm}. However, this fact is not essential to the purpose of the present paper. }  Lagrangian describing the RSB sector:
\begin{equation}
    \label{eq:noscaleL}
    \Lag^{\rm ns}_{\rm matter}\equiv-\frac{1}{4}F_{\mu \nu}^MF^{\mu \nu}_M+ \frac{1}{2}(D_\mu \phi)_a (D^\mu \phi)_a + \bar{\psi}_ji\slashed{D}\psi_j - \frac{1}{2}(Y_{ij}^a \psi_i\psi_j\phi_a+ \text{h.c.})-V_{\rm ns}(\phi).
\end{equation}
We take into account generic numbers 
of real scalars $\phi_a$,   Weyl fermions $\psi_j$ and vectors $V^A_\mu$ (with field strength $F_{\mu\nu}^A$), respectively. The gauge fields $V^A_\mu$ allow us to construct the covariant derivatives \begin{equation}
    \label{eq:covD}
D_\mu\phi_a=\partial_\mu\phi_a+i\theta_{ab}^MV_\mu^M\phi_b,\hspace{1cm} D_\mu\psi_j=\partial_\mu\psi_j+it_{jk}^MV_\mu^M\psi_k,
\end{equation}
where $\theta^M$ and $t^M$ are the generators of the gauge group in the scalar and fermion representations. We include the gauge couplings in the definition of the generators. Note that since all the scalars are real, the Hermitian matrices $\theta^M$ are all purely imaginary and antisymmetric. All indices in (\ref{eq:noscaleL}) are contracted in a gauge-invariant way. 
Also, the $Y^a_{ij}$  are the Yukawa couplings, which are symmetric in $i\leftrightarrow j$, and
\begin{equation}
    \label{eq:noscaleV}
    V_{\rm ns}(\phi) = \frac{\lambda_{abcd}}{4!}\phi_a\phi_b\phi_c\phi_d
\end{equation}
is the no-scale scalar potential, where the quartic couplings $\lambda_{abcd}$ are symmetric in any exchange of the indices.  

In the RSB mechanism the scalar potential develops a flat direction at a certain renormalization scale $\mu=\tilde \mu$. Such direction is parametrized by the scalar field $\chi$  as $\phi_a=\chi\nu_a$ (where $\nu$ is a unit vector in the scalar-field space, $\nu_a\nu_a=1$). The RG-improved potential $V$ along $\nu$ is
\begin{equation}
   V(\chi)=\frac{\lambda_\chi(\mu)}{4}\chi^4,\hspace{1cm}
    \lambda_\chi(\mu)\equiv\frac{1}{3!}\lambda_{abcd}(\mu) \nu_a\nu_b\nu_c\nu_d,
\end{equation}
The requirements that the flat direction occurs at $\mu=\tilde\mu$ and that such direction corresponds to minima of the potential lead, respectively, to the conditions
\begin{equation}
\label{eq:GWcondition}
    \lambda_\chi(\tilde \mu)=0, \hspace{1 cm} \lambda_{abcd}(\tilde \mu)\nu^b\nu^c\nu^d =0.
\end{equation}
The one-loop quantum effective potential renormalized at $\mu=\tilde\mu$ is 
\begin{equation}
\label{eq:1loopP}
    V_q(\chi)= \frac{\bar{\beta}}{4}\left(\text{log}\frac{\chi}{\chi_0}-\frac{1}{4}\right)\chi^4,
\end{equation}
where 
\begin{equation}
    \bar{\beta}\equiv \left[\mu \frac{d\lambda_\chi}{d\mu}\right]_{\mu=\tilde{\mu}}
\end{equation}
and $\chi_0$ is the scale introduced via dimensional transmutation by the RSB mechanism and related to $\tilde{\mu}$ through the renormalization-scheme-dependent formula. We require $\bar \beta>0$ to ensure that the quantum one-loop potential along the flat direction has the absolute minimum for $\chi=\chi_0$. The flat-direction field $\chi$ radiatively develops a mass, which is given by 
\begin{equation}
    \label{eq:masschi}
    m_\chi= \sqrt{\bar{\beta}}\chi_0.
\end{equation}

\section{Interactions}\label{Interactions}
We are interested in the fluctuations around the true vacuum, $\delta \chi \equiv \chi-\chi_0$, and the relative decay channels. In this section we study the leading interactions responsible for these channels.

\subsection{Scalar and vector interactions}

Interactions between scalars and vectors come from the covariant kinetic term 
\begin{equation}
\label{eq:Dterm}
    \frac{1}{2}(D_\mu \phi)_a(D^\mu\phi)_a=\frac{1}{2}\partial_\mu\phi_a\partial^\mu \phi_a+i\theta_{ab}^MV_\mu^M\phi_b\partial^\mu\phi_a+\frac{1}{2}\theta_{ba}^M\theta_{ac}^NV_\mu^MV^{N\mu}\phi_b\phi_c,
\end{equation}
We choose the unitary gauge \cite{Weinberg:1996kr}: since it is always possible  to rotate the  scalar fields in such a way that the $\phi_a$ are orthogonal to the would-be Goldstone directions (in the scalar-field space)\footnote{This is true for compact groups.}, if any, we can impose the gauge condition
\begin{equation}
    \label{eq:UnitaryGauge}
    \phi_a\theta_{ab}^N \nu_b=0,\hspace{1cm} \text{for all $N$}.
\end{equation}
Note that $\theta_{ab}^N \nu_b\neq 0$ when $\theta^N$ is broken.
 
The last term in (\ref{eq:Dterm}) gives the vector mass matrix 
\begin{equation}
    (M^2_V)_{MN}\equiv \nu^b\theta^M_{ba}\theta^N_{ac}\nu^c\chi_0^2.
\end{equation}
The diagonalization of the corresponding matrix $M_V^2$   can be obtained redefining the generators as
\begin{equation}
    \theta'^M= \mathcal{M}_{MN}\theta^N,
\end{equation}
 for some coefficients $\mathcal{M}_{MN}$. 
 After this is done, $M_V^2$ is diagonal and reads  
\begin{equation}
\label{eq:MVdiag}
    (M_V^2)_{MN}= - T_MT_N\chi_0^2,
\end{equation}
where we defined the ``projection" of the generator along the flat direction
\begin{equation}
\label{eq:Proj}
    T^M_a\equiv \theta'^M_{ab}\nu_b=\begin{cases}0 &\theta'^M\,\text{ is unbroken}\\ \ne 0 & \theta'^M\, \text{is broken}  \end{cases}
\end{equation}
We call $B^M_\mu$ the massive vectors corresponding to the broken $\theta'^M$ and we call $A^M_\mu$ the massless vectors corresponding to the unbroken $\theta'^M$.

It is then convenient to decompose the scalar fields as follows
\begin{equation}
    \phi_a=\phi_a^\bot+\chi\nu_a, 
\end{equation}
with $\phi_a^\bot$ satisfying the condition $\phi_a^\bot\nu_a=0$.
Then the classical mass terms of the scalar fields read
\begin{equation}
    \frac{\chi_0^2}{2}\lambda_{abcd}\nu_c \nu_d \phi_a \phi_b= \frac{\chi_0^2}{2}\lambda_{abcd}\nu_c\nu_d\phi_a^\bot\phi_b^\bot+\chi_0^2\lambda_{abcd}\nu_b \nu_c \nu_d \phi^\bot_a\chi+\frac{\chi_0^2}{2}\lambda_{abcd}\nu_a\nu_b\nu_c\nu_d\chi^2.
\end{equation}
At the energy scale of interest, $\mu=\tilde\mu$, the last two terms are zero because of \eqref{eq:GWcondition}. Thus,  to diagonalize the scalar mass matrix we can act with a real orthogonal matrix $\mathcal{O}$ on the fields $\phi^\bot_a$ only,
\begin{equation}
    \phi_a'=\mathcal{O}_{ab}\phi_b^\bot.
\end{equation}
Since the matrix $\mathcal{O}$ acts only on the space orthogonal to the flat direction, $\phi_a'\nu_a=0$. After this diagonalization is performed, the diagonal mass term reads
\begin{equation}
    \frac{1}{2}\lambda_a\chi^2_0\phi_a'\phi_a',
\end{equation}
where the $\lambda_a$ are some real coefficients. Also, the $\lambda_a$ are always non negative for all theories with Lagrangian of the form in~(\ref{eq:noscaleL})~\cite{Salvio:2023qgb}. 
At the same time $\delta \chi$ acquires mass only radiatively, Eq.~(\ref{eq:masschi}). 
  
As a result, the interaction Lagrangian in unitary gauge that is relevant for the $\delta \chi$ decay  is 
\begin{equation}
    \label{eq:LVint}
    \Lag^V_{\rm int}= -(\bar{T}^N)^2 \chi_0 \delta\chi B_\mu^N B^{N\mu}-\bar{T}^N_a\bar{\theta'}_{an}^N\mathcal{O}_{nc}^T\delta\chi \phi_c' B_\mu^N B^{N\mu},
\end{equation}
where $\bar{T}^N_a\equiv\bar \theta'^N_{ab}\nu_b$ is the ``projection" along the flat direction of the broken generators, which we henceforth call $\bar \theta'^N$. Also $(\bar{T}^N)^2\equiv\bar{T}^N_a\bar T^N_a$, where, while the index $a$ is summed, $N$ is not.  There is no interaction between $\delta\chi$ and the massless gauge bosons $A_\mu^N$ at tree level because the ``projection" of unbroken generators along the flat direction is always zero. The terms 
\begin{equation}
    -i\bar{T}^N_a \partial^\mu\delta\chi B_\mu^N\phi_a^\bot + i \bar{T}^N_a\delta\chi B_\mu^N\partial^\mu\phi_a^\bot  + i\bar{T}^N_a\chi_0  B_\mu^N\partial^\mu\phi_a^\bot-i\bar{T}^N_a \partial^\mu\delta\chi B_\mu^N\chi_0 \nu_a 
\end{equation}
do not appear in the interaction Lagrangian \eqref{eq:LVint} because the first three terms are zero in the unitary gauge and the last term is zero in every gauge, because of the anti-symmetry of the generators.

Moreover, the decays of $\delta\chi$ into scalar fields are described by the Lagrangian
\begin{equation}
\label{eq:Sint}
    \Lag_{\rm int}^S= -\lambda_{a} \chi_0\delta\chi  \phi_a'\phi_a'-\frac{1}{3!}\lambda_{mnl}\mathcal{O}_{ma}^T\mathcal{O}^T_{nb}\mathcal{O}^T_{lc}\delta\chi \phi_a' \phi_b' \phi_c'.
\end{equation}
where
\begin{equation}
     \lambda_{mnl}\equiv \lambda_{mnld}(\tilde{\mu})\nu^d.
\end{equation}

\subsection{Fermion interactions}\label{Fermion interactions}
The fermion mass matrix is related to the term 
\begin{equation}
   \frac{1}{2}(Y^a_{ij} \psi_i\psi_j\phi_a+ \text{h.c.}).
\end{equation}
We can choose a fermion  basis such that $\mu_F \equiv Y^a\nu_a\chi_0$ (as well as $\mu_F^\dagger$) is diagonal and the (diagonal) square mass matrix is~\cite{Salvio:2023qgb}
\begin{equation}
M^2_F\equiv \mu_F\mu_F^\dagger=Y_\nu Y_\nu^\dagger \chi_0^2\equiv \text{diag}(\dots,|y_i|^2\chi_0^2,\dots),\hspace{1cm}Y_\nu\equiv Y^a \nu_a,
\end{equation}
where the $Y^a$ are the Yukawa matrices with elements $Y^a_{ij}$.
 In this basis also the interaction term with $\delta \chi$ is diagonalized with coupling $y_i$.

\section{Decay Rates}\label{Decay Rates}
We proceed to  calculate the decay rate of $\delta \chi$ in scalar, vectors and fermions. 
First we note that thanks to probability conservation the total (inclusive) decay rate is independent from the choice of basis for the final states. Indeed, given a set of final states $|\alpha_i\rangle$ and a unitary transformation that changes the basis of the final states 
\begin{equation}
    |\beta_i\rangle = U_{ij}|\alpha_j\rangle,
\end{equation}
the inclusive decay rate of a system in the initial state $|\delta \chi\rangle$ into the final states $|\beta_j\rangle$ is described by $\sum_j|\langle\beta_j|S|\delta \chi\rangle|^2$, where $S$ is the scattering operator (for the following proof, we explicit show the sum over the indices). Using the unitary condition $\sum_j U^*_{ji} U_{jk}= \delta_{ik}$ we obtain 
\begin{equation}
\begin{aligned}
\sum_j |\langle \beta_j | S | \delta \chi \rangle|^2 
&= \sum_j \langle \beta_j | S | \delta \chi \rangle \langle \delta \chi | S^\dagger | \beta_j \rangle 
= \sum_{ijk} U^*_{ji} U_{jk} \langle \alpha_i | S | \delta \chi \rangle \langle \delta \chi | S^\dagger | \alpha_k \rangle \\
&= \sum_{ik} \delta_{ik} \langle \alpha_i | S | \delta \chi \rangle \langle \delta \chi | S^\dagger | \alpha_k \rangle 
= \sum_i |\langle \alpha_i | S | \delta \chi \rangle|^2.
\end{aligned}
\end{equation}
Thus we can calculate inclusive decay rates without worrying about the basis of the final states.

The two-body decay rates into two scalars, $ \Gamma^{(2S)}$, two fermions, $ \Gamma^{(2F)}$ and two vectors, $ \Gamma^{(2V)}$,  are:
\begin{equation}
    \Gamma^{(2S)}=\sum_a\frac{\lambda_a^2\chi_0}{8\pi \sqrt{\bar \beta}}\sqrt{1-\frac{4\lambda_a}{\bar \beta}} \,\Theta\left(\sqrt{\bar \beta}-2\sqrt{\lambda_i}\right)\equiv \sum_a\Gamma(\delta \chi\rightarrow \phi'_a\phi'_a),
\end{equation}
\begin{equation}
\label{eq:ferm2}
    \Gamma^{(2F)}= \sum_i\frac{s_iy_i^2\chi_0}{8\pi}\sqrt{\bar\beta}\left(1-\frac{4y_i^2}{\bar \beta}\right)^{3/2} \,\Theta\left(\sqrt{\bar \beta}-2y_i\right)\equiv \sum_i\Gamma(\delta\chi\rightarrow \psi_i' \psi_i'),
\end{equation}
\begin{equation}
\label{eq:vet2}
    \Gamma^{(2V)}=\frac{g^4_N \chi_0}{32\pi \sqrt{\bar \beta}}\sum_N\left(12+\frac{\bar \beta^2}{g_N^4}-4\frac{\bar \beta}{g_N^2}\right)\sqrt{1-\frac{4g_N^2}{\bar \beta}}\,\Theta\left(\sqrt{\bar \beta}-2g_N\right)\equiv \sum_N \Gamma(\delta \chi\rightarrow B_N B_N),
\end{equation}
where $\Theta$ is the Heaviside step function, $\psi'_i$ are fermions in the mass basis defined in Sec.~\ref{Fermion interactions}, $s_i$ is a symmetry factor that is 1 for Dirac fermions and 1/2 for Majorana fermions and  
 $g_N\equiv\sqrt{-(\bar T^N)^2}$, which is real and positive as the $\bar T_N$ are purely imaginary.  In~\cite{Djouadi:2005gi1,Djouadi:2005gj2} analogous calculations are performed for the SM Higgs physics and in the Minimal Supersymmetric Standard Model.

In general we also have three-body decay processes ($\delta \chi \rightarrow \phi'_a\phi'_b\phi'_c$ and $\delta \chi \rightarrow \phi'_aB_NB_N$):
\begin{equation}
\Gamma^{(3S)}\equiv\sum_{a,b,c} \Gamma(\delta \chi\rightarrow \phi'_a\phi'_b\phi'_c), \qquad \Gamma^{(SV)} 
    \equiv \sum_{c,N} \Gamma(\delta \chi \rightarrow \phi'_c B^N B^N).
\end{equation}
These are explicitly given by
\bea
\Gamma^{(3S)} =\sum_{a,b,c}\frac{S\chi_0}{64\pi^5\sqrt{\bar \beta}}(\lambda_{abc}')^2\omega_{3}\left(\bar \beta,\sqrt{\lambda_a},\sqrt{\lambda_b},\sqrt{\lambda_c}\right)\Theta\left(\sqrt{\bar \beta}-\sqrt{\lambda_a}-\sqrt{\lambda_b}-\sqrt{\lambda_c}\right),
\nonumber \\   
    \Gamma^{(SV)} = \sum_{N,c} \frac{(G_c^N)^4 \chi_0}{32\pi^5\sqrt{\bar \beta}} \left( 
        2\omega_3\left(\bar \beta, g_N, g_N, \sqrt{\lambda_c}\right) 
        + \frac{1}{g_N^4}\omega_3^{(4)}\left(\bar \beta, g_N, \sqrt{\lambda_c}\right) 
    \right)   \Theta\left(\sqrt{\bar \beta} - \sqrt{\lambda_c} - 2g_N\right)\nonumber,
\eea
where $S=1/n_f!$ is a symmetry factor that depends on the number $n_f $ of identical particles in the final state, $\lambda_{abc}'\equiv \lambda_{mnl}\mathcal{O}^T_{ma}\mathcal{O}^T_{nb}\mathcal{O}^T_{lc}$. Moreover, $(G_c^N)^4\equiv  (\bar T_a^N \bar {\theta}'^{N}_{an}\mathcal{O}_{nc}^T)^2$,  where we do not sum over the repeated index $N$.
    Also, the three-body phase space integrals $\omega_3$ and $\omega_3^{(4)}$ are calculated in Appendix \ref{appendix:3ps}.

The three-body decays are not always negligible compared to the two-body ones: if the couplings $G_c^N$ and $\lambda'_{abc}$ are large with respect to the couplings of the two-body decay, $\lambda_a$, $g_N$, $y_i$, then the contribution of three-body decay processes can be comparable to the two-body ones. Moreover, we are interested in the inclusive decay rate and the number of interactions that contribute to  $\Gamma^{(3S)}$ and $\Gamma^{(SV)}$ are $\binom{N_s + 2}{3}$ 
and $N_s N_{bg}$, respectively, where $N_s$ is the number of scalar degrees of freedom and $N_{bg}$ is the number of broken gauge generators. Moreover, each diagram is then multiplied by an appropriate symmetry factor. For two-body decays, instead, the number of interactions is proportional to the number of vectors, scalars and fermions  coupled to $\delta \chi$, for vector, scalar, and fermion decays, respectively.  Since the number of scalar degrees of freedom can easily reach $N_s\sim 10$, the contributions  of three-body decays to inclusive decay rates may be comparable to the contributions of two-body decays.

However, there are cases where we can neglect the three-body decays: one is  when the $G_c^N$ and $\lambda_{abc}'$ are small  and another one is when  the gauge symmetry of the specific model under study forbids the presence of the corresponding interactions.

To study the case where three-body decays are negligible  we define the following parameters:
\begin{equation}
    \zeta_{Sa}\equiv\frac{4m_{a}^2}{m^2_\chi}= \frac{4\lambda_a}{\bar \beta}, \hspace{1cm}
    \zeta_{Fi}\equiv \frac{4m_i^2}{m_\chi^2}=\frac{4y^2_i}{\bar \beta}, \hspace{1cm}
    \zeta_{VN}\equiv \frac{4m_N^2}{m_\chi^2}=\frac{4g^2_N}{\bar \beta}.  \label{zeta1}
\end{equation}
These parameters measure how small the masses of the products with respect to the mass of $\delta \chi$ are.  Note that  when one of these $\zeta$s are less than 1,  the corresponding decay channel is open. Using these parameters, we can write the decay rates as
\begin{equation}\label{eq:zetadc}
    \Gamma^{(2S)}(\bar \beta,\zeta_S)=\frac{\bar \beta^{3/2}}{32\pi}\zeta_S\chi_0,\hspace{1cm}
    \Gamma^{(2F)}(\bar \beta,\zeta_F)=\frac{\bar \beta^{3/2}}{32\pi}\zeta_F\chi_0,\hspace{1cm}
    \Gamma^{(2V)}(\bar \beta,\zeta_V)=\frac{\bar \beta^{3/2}}{32\pi}\zeta_V\chi_0,
\end{equation}
where we defined the parameters
\begin{align}
    \zeta_S&\equiv \frac{1}{4}\sum_a\zeta_{Sa}^2\sqrt{1-\zeta_{Sa}}\Theta\left(1- \zeta_{Sa}\right), \nonumber  \\
     \zeta_F&\equiv\sum_i s_i\zeta_{Fi}(1-\zeta_{Fi})^{3/2}\Theta\left(1- \zeta_{Fi}\right), \label{eq:Gzdc}\\
    \zeta_V&\equiv \sum_N\left(1-\zeta_{VN}+\frac{3}{4}\zeta_{VN}^2\right)\sqrt{1-\zeta_{VN}}\Theta\left(1-\zeta_{VN}\right). \nonumber
\end{align}
We get these simple expressions  because the couplings that regulate the interactions are the same couplings that determine the masses of the particles. 
Putting all together we can write the total inclusive two-body decay rate as follows:
\begin{equation}
    \Gamma_{\rm tot}(\bar \beta, \zeta_S,\zeta_F, \zeta_V)\equiv\Gamma^{(2S)}(\bar \beta,\zeta_S)+\Gamma^{(2F)}(\bar \beta, \zeta_F)+\Gamma^{(2V)}(\bar \beta, \zeta_V)= \frac{\bar \beta^{3/2}}{32\pi}\zeta_{\rm tot}\chi_0,
\end{equation}
where     $\zeta_{\rm tot}= \zeta_S+\zeta_F+\zeta_V$.

We can study the limit when the $\zeta$s in~(\ref{zeta1}) are all small, which corresponds to the case when all the products are much lighter than $\delta \chi$.
 In this case we get 
    \begin{align}
    \label{eq:smallZ}
        \zeta_S&= \frac{1}{4}\sum_a\zeta_{Sa}^2+ O(\zeta_{Sa}^3),\nonumber\\
        \zeta_F&=\sum_is_i\left(\zeta_{Fi}-\frac{3}{2}\zeta_{Fi}^2\right)+O(\zeta_{Fi}^3),\\
        \zeta_V&=N_{bg}- \sum_N\left(\frac{3}{2}\zeta_{VN}-
        \frac{9}{8}\zeta_{VN}^2\right)+O(\zeta_{VN}^3).\nonumber
    \end{align}
Note that in the last expression there is a contribution  $N_{bg}$, which survives in the $g_N\to 0$ limit.  This is due to the fact that the Goldstone bosons do not decouple  in such limit.

\section{Radiative electroweak symmetry breaking} \label{sec:RSBEW}

Before applying the previous results to the calculation of reheating it is necessary to explain how the energy stored in $\delta\chi$ can be transferred to SM particles. Indeed, for successful reheating the SM particles should be brought to thermal equilibrium at the reheating temperature. 

To this purpose we need to explain how EWSB can be related to RSB. Efficient reheating occurs when the RSB sector is not a hidden dark sector, but includes the SM.  
In this case it is possible to generate the EW scale through  RSB  via the coupling between the field $\mathcal{H}$, i.e.~the complex scalar doublet of the SM, and the RSB sector. Writing  $\mathcal{H}$  as 
\begin{equation}
    \mathcal{H}=\frac{1}{\sqrt{2}}\begin{pmatrix}\eta_1+i\eta_2\\ h+i\eta_3\end{pmatrix}
\end{equation}
where the $\eta$ fields are the three  SM would-be Goldstone bosons, which appear only as longitudinal degrees of freedom of the vector bosons in the gauge  we adopt (the unitary  gauge). So, the tachyonic mass term of the field $h$ can be generated by
\begin{equation}
    \Lag_{\phi h}=\frac12 \lambda_{ab} \phi_a\phi_b |{\mathcal H}|^2,\label{Hportal}
\end{equation}
where the $\lambda_{ab}$ are some of the quartic couplings. 
The term above can generate the electroweak scale radiatively~\cite{Salvio:2023qgb} in a generic RSB theory. 

For the sake of clarity, however,  let us now restrict  ourselves to the case where the only relevant physical scalars in the theory are $h$ and a real scalar $\phi_h$ and assume a $\mathbb{Z}_2$ symmetry such that $\phi_h\rightarrow -\phi_h$. Further generalizations are straightforward. In this case the classical scalar potential has the form (see \cite{Hambye:2013dgv,DelleRose:2019pgi,Liu:2024fly} for similar models):
\begin{align}
V\left(h,\phi_h\right)& = \frac{1}{4}\lambda_hh^4 + \frac{1}{4}\lambda_\phi\phi_h^4 - \frac{1}{4}\lambda_{\phi h}h^2\phi_h^2 \\&= \frac{1}{4}\left(\sqrt{\lambda_h}h^2-\sqrt{\lambda_\phi}\phi_h^2\right)^2+\frac{1}{2}\sqrt{\lambda_h}\sqrt{\lambda_\phi}h^2\phi_h^2-\frac{1}{4}\lambda_{\phi h}h^2\phi_h^2 \nonumber ,
\end{align}
where $\lambda_h$, $\lambda_\phi$ and $\lambda_{\phi h}$ are the relevant quartic couplings.

At the renormalization scale $\tilde \mu$ the classical potential is zero along the flat direction, and  couplings and fields satisfy
\begin{equation}
   \sqrt{\lambda_h}h^2=\sqrt{\lambda_\phi}\phi_h^2~~\mbox{(on the flat direction)}, \hspace{1cm} \lambda_{\phi h}= 2\sqrt{\lambda_h}\sqrt{\lambda_\phi}. \label{FDsimple}
\end{equation}
Henceforth in this section,  we always consider the theory renormalized at $\tilde\mu$.
The first relation in~(\ref{FDsimple}) means that we can rotate the fields in a way that the flat-direction field $\chi$ manifestly appear in the Lagrangian
\begin{equation}
    \begin{cases}\label{eq:rotation}
    \phi_h=\chi \cos \alpha- H\sin \alpha,\\ 
    h=\chi \sin \alpha+H \cos \alpha,
    \end{cases}
\end{equation}
where the mixing angle $\alpha$ is defined here by
\begin{equation}
    \tan \alpha\equiv\sqrt{\frac{\lambda_{\phi h}}{2\lambda_h}},
\end{equation}
and $H$ is the field of the (observed) Higgs boson.
Including the 1-loop contribution of the potential along the flat direction, given in~\eqref{eq:1loopP}, and expanding the flat-direction field as $\chi=\chi_0+ \delta \chi$, the relation between the RSB scale and the SM vacuum expectation value (VEV) $v$ is
\begin{equation}
\label{eq:vev}
    v=\chi_0 \sin \alpha= \chi_0\sqrt{\frac{\lambda_{\phi h}}{\lambda_{\phi h}+2\lambda_h}}.
\end{equation}
The second equation in \eqref{eq:rotation} then reads $h=v+ \delta \chi \sin \alpha +H \cos \alpha$.
After this field redefinition the effective 1-loop potential reads 
\begin{align}
    V_{\rm 1-loop}(H,\delta \chi)=&\frac{\lambda_{\phi h}\chi_0^2}{2}H^2 +\frac{4
\lambda_{ h}^2-\lambda_{\phi h}^2}{2\sqrt{2} \lambda_h}\sqrt{\frac{\lambda_{\phi h}\lambda_h}{(\lambda_{\phi h}+2\lambda_h)^2}}\, \chi_0H^3+\frac{(\lambda_{\phi h}-2\lambda_h)^2}{16 \lambda_h}H^4 \nonumber\\ 
    &+ \lambda_{\phi h} \chi_0\delta \chi H^2+\frac{\lambda_{\phi h}}{2}\delta \chi^2 H^2+ \frac{4\lambda_{ h}^2-\lambda_{\phi h}^2}{2\sqrt{2} \lambda_h}\sqrt{\frac{\lambda_{\phi h}\lambda_h}{(\lambda_{\phi h}+2\lambda_h)^2}}\, \delta \chi H^3 +\frac{\bar \beta}{2}\chi_0^2 \delta \chi^2\\ \nonumber
    &+ \delta\chi \text{ self-interactions}.
\end{align}
The mass of the Higgs boson is
    $M_h=\sqrt{\lambda_{\phi h}}\,\chi_0$. 
Also, the EW gauge bosons  and SM fermions develop a coupling with $\delta \chi$. For example, for the $W$ bosons the interaction Lagrangian with $h$ is given by
\begin{align}
    \Lag_{hWW}= &\frac{1}{4}g_2^2h^2 W^+_\mu W^{-\mu}\nonumber \\ =& \frac{1}{4}g_2^2\chi_0^2\sin^2\alpha \,W^+_\mu W^{-\mu}+\frac{1}{4}g_2^2 H^2  W^+_\mu W^{-\mu} \cos^2\alpha +\frac{1}{2}g_2^2\chi_0\cos \alpha\, \sin \alpha\, H W^+_\mu W^{-\mu}\\ &+\frac{1}{2}g_2^2\chi_0 \sin^2 \alpha \,\delta \chi  W^+_\mu W^{-\mu}+ \frac{1}{4}g_2^2 \sin^2 \alpha \, \delta \chi^2 W^+_\mu W^{-\mu} + \frac{1}{2}g_2^2 \cos \alpha \, \sin \alpha \, \delta \chi H W^+_\mu W^{-\mu}, \nonumber
\end{align}
where $g_1$ and $g_2$ are the gauge constants of the $SU(2)$ and $U(1)$ SM gauge-group factors.  We correctly reproduce the SM terms, such as the mass term for the $W$ boson
\begin{equation}
    M_W= \frac{1}{2}g_2\chi_0 \sin\alpha= \frac{1}{2}g_2 v, 
\end{equation}
plus additional interaction terms with $\delta \chi$ suppressed by some positive power of  $\sin \alpha$. 
The same happens for the $Z$ boson, its  interaction Lagrangian with $h$ is given by
\begin{align}
     \Lag_{hZZ}= &\frac{1}{8}g_Z^2  h^2 Z_\mu Z^\mu \nonumber \\ =& \frac{1}{8}g_Z^2  \chi_0^2\sin^2\alpha \,Z_\mu Z^\mu +\frac{1}{8}g_Z^2   H^2  Z_\mu Z^\mu  \cos^2\alpha +\frac{1}{4}g_Z^2  \chi_0\cos \alpha\, \sin \alpha\, HZ_\mu Z^\mu \\ &+\frac{1}{4}g_Z^2  \chi_0 \sin^2 \alpha \,\delta \chi Z_\mu Z^\mu+ \frac{1}{8}g_Z^2 \sin^2 \alpha \, \delta \chi ^2Z_\mu Z^\mu + \frac{1}{4}g_Z^2 \cos \alpha \, \sin \alpha \, \delta \chi H Z_\mu Z^\mu \nonumber,
\end{align}
where 
\begin{equation}
    M_Z=\frac{1}{2}g_Z\, \chi_0 \sin \alpha=\frac{1}{2}g_Z\, v,\hspace{1cm} g_Z\equiv \sqrt{g_1^2+g_2^2}.
\end{equation}

\begin{figure}[t!]
        \includegraphics[scale=0.50]{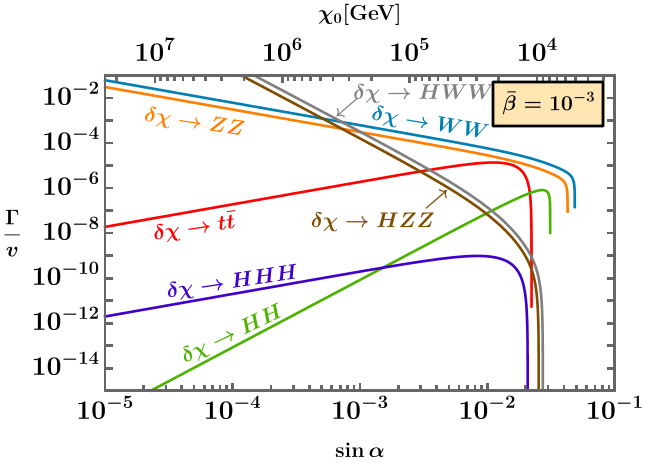} 
~ \includegraphics[scale=0.50]{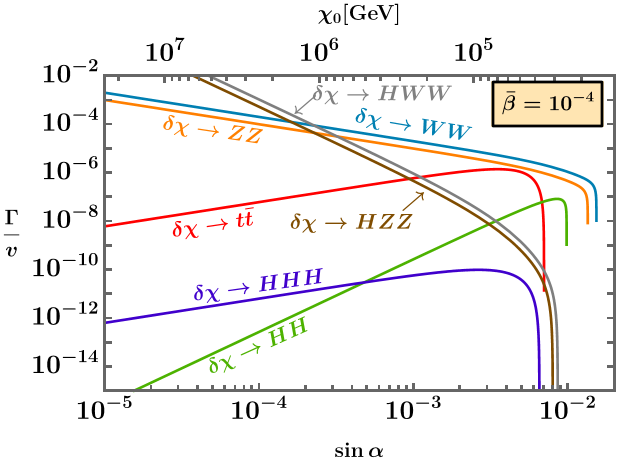}
    \caption{\it Relevant decay rates, $\Gamma$, in units of the SM VEV $v$, of $\delta \chi$ into SM particles as a function of the mixing angle $\alpha$ or, equivalently, the radiative symmetry breaking scale $\chi_0$.}
    \label{fig:DRSMbeta-4}
\end{figure}

The mechanism works in a similar  way for the SM fermions, for which the Yukawa interactions with $\delta \chi$ are given by the SM Yukawa interactions with the  Higgs field, but, unlike for the Higgs field, with a $\sin \alpha$ suppression. The decay rate of $\delta \chi$ into SM particles is shown in Fig.~\ref{fig:DRSMbeta-4}. The mixing angle $\alpha$ should be small to respect the experimental constraints, so in order to generate the SM VEV $v$, $\chi_0\gg v$. In this limit we can approximate 
$$\bar\beta\approx\left[\mu \frac{d\lambda_{\phi}}{d\mu}\right]_{\mu=\tilde{\mu}}.$$

Another useful way to describe the couplings of $\delta \chi$  to the SM massive gauge bosons is  to consider the fact that the flat direction, the $\nu_a$,  depends on the mixing angle $\alpha$ and
\begin{equation}
    \bar T_a^N(\alpha)\equiv \bar \theta'^N_{ac} \nu_c(\alpha),
     \hspace{1cm}\nu(\alpha)\equiv (\sin \alpha, \cos \alpha, 0, 0, ...),
\end{equation}
where the zero entries in $\nu(\alpha)$ refer to possible (would-be) Goldstone directions. 
The fields $h$ and $\phi_h$ are parametrized along the flat direction $\chi$ as
\begin{equation}
    \phi_h= \chi \cos \alpha,\hspace{1cm} h=\chi \sin \alpha,
\end{equation}
so we get the expression
\begin{equation}
    \nu(\alpha)_c\bar \theta'^N_{ca}\bar \theta'^N_{ab}\nu_b(\alpha)\chi^2B_\mu^NB^{\mu N}=\bar \theta'^{N_{\rm SM}}_{1a}\bar \theta'^{N_{\rm SM}}_{a1}\sin^2\alpha \,\chi^2 B^{N_{\rm SM}}_\mu B^{\mu N_{\rm SM}}+\bar \theta'^{N_{\rm D}}_{2a}\bar \theta'^{N_{\rm D}}_{a2}\cos^2\alpha\, \chi^2B^{N_D}_\mu B^{\mu N_D}, \nonumber
\end{equation}
where $N_{\rm SM}$ is the index $N$ running only on SM massive gauge bosons and $N_{\rm D}$ is the index of the dark-sector massive gauge bosons, if any ($\mathcal{H}$ is assumed to be neutral with respect to a possible extra gauge factor of the dark sector). Note that the term proportional to $\sin \alpha \cos \alpha$ is not present because $\phi_h$ is assumed to be a  singlet under the SM gauge group.

To understand  the decay rates in Fig.~\ref{fig:DRSMbeta-4} it is useful to analyze their behavior for small values of $\sin \alpha$: since the $\zeta$s in~(\ref{zeta1}) scale in general like $\zeta\sim \sin^2 \alpha$  at $v$ and SM masses fixed, using \eqref{eq:zetadc} and \eqref{eq:Gzdc}, we get for the two body decays,
\begin{equation}
	\frac{\Gamma_{VV}}{v}\sim \frac{1}{\sin \alpha}, \quad 
	\frac{\Gamma_{HH}}{v}\sim  \sin^3 \alpha, \quad 
	\frac{\Gamma_{f\bar f}}{v}\sim  \sin \alpha.
\end{equation}
Notice that, while the decay rate into scalars, $\Gamma _{HH}$, and fermions, $\Gamma _{f\bar f}$,  increase with $\sin \alpha$, the decay rate into vectors, $\Gamma _{VV}$, decreases. In order to consider the scaling of the three-body decays we notice that 
\begin{equation}
	\lambda_{\phi h}\sim \sin^2 \alpha, \hspace{1cm} \lambda_{HHH}\sim \sin\alpha,\hspace{1cm} G^2\sim \sin \alpha,
\end{equation}
where $\lambda_{HHH}$ and $G^2$ are the couplings in the Lagrangian that regulate the three-body decays into  three scalars and in two vectors and one scalar, respectively. From \eqref{eq:ODim} we see that 
\begin{equation}
	\Omega_3\sim \sin^{-2} \alpha,\hspace{1cm} \Omega_3^{(4)}\sim \sin^{-6} \alpha
\end{equation}
and finally the three-body decay rates into three Higgs bosons and into one Higgs boson and two identical vectors scale as, respectively,
\begin{equation}
	\frac{\Gamma_{HHH}}{v}\sim \frac{\lambda_{HHH}^2\sin \alpha}{v^2}\Omega_3\sim \sin^3\alpha \cdot\sin^{-2} \alpha= \sin \alpha,
\end{equation}
\begin{equation}
	\frac{\Gamma_{HVV}}{v}\sim \frac{G^4\sin \alpha}{v^2}\left(2\Omega_3+\frac{1}{m_V^4}\Omega_3^{(4)}\right)\sim \sin^3\alpha \cdot \sin^{-6}\alpha=\sin^{-3}\alpha,
\end{equation}
where $m_V$ is the vector mass. From this scaling laws for small $\alpha$ it is clear that $\Gamma_{HVV}$ grows much faster than $\Gamma_{VV}$ as $\alpha$ decreases, explaining the corresponding behaviors in Fig.~\ref{fig:DRSMbeta-4}.

\section{Reheating through the decays of the flat-direction field}\label{Reheating}

If $\delta\chi$ decays into some SM particles with width $\Gamma_{\delta\chi}$ the reheating temperature $T_{\rm rh}$ is at least
\be T_{\rm rh}\gtrsim \min\left( \left(\frac{45 \Gamma^2_{\delta\chi}\bp^2}{4\pi^3 g_*(T_{\rm rh})}\right)^{1/4},\left(\frac{30 \Delta V}{\pi^2 g_*(T_{\rm rh})}\right)^{1/4}\right),\label{Trdef}\ee
where $\bar M_P$ is the reduced Planck mass, $g_*(T)$ is the effective number of relativistic species in thermal equilibrium at temperature $T$ and $\Delta V$ is the vacuum energy density due to $\delta\chi$,   
 \be \Delta V = \frac{\bar\beta \chi_0^4}{16}.\label{DeltaV} \ee
Note that $\Delta V$ represents the full energy budget of the system. 

The mechanism described in this section is the standard perturbative contribution to reheating. In Sec.~\ref{Preheating} we will also study preheating through parametric resonance.

We can define the equivalence temperature $T_{\rm eq}$ as 
\begin{equation}
   \frac{\pi^2}{30}g_*(T_{\rm eq}) T_{\rm eq}^4\equiv \Delta V,
\end{equation}
which, using~(\ref{DeltaV}), leads to
\begin{equation}
\label{eq:Teq}
    T_{\rm eq}^4= \frac{15\bar \beta \chi_0^4}{8\pi^2g_*(T_{\rm eq})}.
\end{equation}
With this definition reheating can be considered fast when $T_{\rm rh}\approx T_{\rm eq}$.
Therefore, fast reheating occurs when
\begin{equation}
\label{eq:Trhrel1}
    \zeta_{\rm tot}^2\gtrsim \frac{512 \pi^3}{3\bar\beta^2} \frac{\chi_0^2}{\bar M_P^2},
\end{equation}
where we assumed $g_*(T_{\rm rh})\approx g_*(T_{\rm eq})$. Note that in deriving this condition we have only taken into account two-body decays. However, including  the three-body decays, which can easily be done with the results of the previous section, can only make reheating even faster.
When the $\zeta$s in~(\ref{zeta1}) are small, we can approximate $\zeta_{\rm tot}\approx N_{bg}$ and the fast reheating condition in~(\ref{eq:Trhrel1}) becomes
\begin{equation}
\label{eq:Trhrel2}
    N_{bg}^2\gtrsim  \frac{512 \pi^3}{3\bar\beta^2} \frac{\chi_0^2}{\bar M_P^2}.
\end{equation}

The relations \eqref{eq:Trhrel1} and \eqref{eq:Trhrel2} hold when the SM is embedded in the RSB sector such that $\delta \chi$ decays directly into the SM particles or into particles with sizable couplings with the SM. In this case, since  EWSB occurs through RSB (see Sec.~\ref{sec:RSBEW}), Eqs.~\eqref{eq:Trhrel1} and~\eqref{eq:Trhrel2} can be written respectively  as
\begin{equation}
\label{eq:TrhrelSM2}
   \zeta_{\rm tot}^2\sin^2 \alpha\gtrsim  \frac{512 \pi^3}{3\bar\beta^2} \frac{v^2}{\bar M_P^2}, \hspace{1cm} N_{bg}^2\sin^2 \alpha\gtrsim  \frac{512 \pi^3}{3\bar\beta^2} \frac{v^2}{\bar M_P^2}.
\end{equation}
 where here we adopted the model-independent definition $\sin \alpha \equiv v/\chi_0$.  Interestingly, the smaller  $\chi_0$ (or, equivalently, the larger $\alpha$) the weaker the  condition for fast reheating, despite $\Gamma_{\delta\chi}\propto \chi_0$ and the inclusive decay rate decreases when $\chi_0$ decreases: this is because the first entry in the $\min$ in~(\ref{Trdef}) scales with $\chi_0$ as $\sqrt{\chi_0 \bar M_P}$, while the second one scales as  $\chi_0$. 

\subsection{Compatibility between reheating and phase transition}\label{Compatibility between reheating and phase transition}

It is important to keep in mind that the conditions for fast reheating are in turn subject to the condition that the supercooled PT has actually taken place. In particular, one should require the existence of the nucleation temperature $T_n$. 

As explained in~\cite{Salvio:2023qgb,Salvio:2023ext,Banerjee:2024cwv}, the PTs in the RSB scenario can be described (for example, an expression for $T_n$ can be derived) in a model-independent way if enough supercooling occurs, specifically when
\be \epsilon\equiv  \frac{g^4}{6\bar\beta \log\frac{\chi_0}{T_n}}
 \label{CondConv}\ee
 is small enough. Here,  given the definitions in~(\ref{zeta1}), the parameter $g$ is defined by  
 \begin{equation}\label{newg}
    \frac{4g^2}{\bar \beta}=\sum_a \zeta_{Sa}+3\sum_{N}\zeta_{VN}+\sum_{i}\zeta_{Fi}.
\end{equation}
If $\epsilon\ll1$ the nucleation temperature can be computed in a small-$\epsilon$ expansion (``supercool expansion"), which at leading order (LO) gives
\begin{equation}
\label{eq:NT}
    T_n\approx \chi_0\exp \left(\frac{\sqrt{c^2-16a}-c}{8}\right),\hspace{0.5cm} a\equiv \frac{c_3 g}{\sqrt{12}\bar \beta}, \hspace{0.5cm}c\equiv 4\log \frac{4\sqrt{3}\bar M_P}{\sqrt{\bar \beta}\chi_0},
\end{equation}
where $c_3=18.8973...$\,.
The expression of $T_n$ in  \eqref{eq:NT} is real only when $c^2\ge16a$; if this condition is not satisfied there is no acceptable solution for $T_n$ in the supercool expansion at LO.  This gives a constraint on $\chi_0$, $\bar\beta$ and $g$ and, thus, on the $\zeta$s:
\begin{equation}
\label{eq:Tn}
    \sqrt{\sum_a\zeta_{Sa}+3 \sum_N \zeta_{VN}+\sum_i\zeta_{Fi}}\le \frac{4\sqrt{3\bar \beta}}{c_3}\log ^2\left(\frac{4\sqrt{3}\bar M_P}{\sqrt{\bar \beta}\chi_0}\right).
\end{equation}

As discussed in~\cite{Salvio:2023ext}, the validity of the supercool expansion can be extended to models with $\epsilon$ at most of order-one  by taking into account the effect of the extra parameter $\tilde g$, which, given the definitions in~(\ref{zeta1}), is defined by
\begin{equation}\label{newgtilde}
    \left(\sqrt{\frac{4}{\bar \beta}}\,\tilde{g}\right)^3=\sum_a\zeta_{Sa}^{3/2}+3\sum_N\zeta_{VN}^{3/2}.
\end{equation}
In this more general case, $T_n$ has been approximated numerically in~\cite{Salvio:2023ext,dataset} for all models of this sort.  Therefore, the requirement of the existence of the nucleation temperature $T_n$ can be obtained by combining~(\ref{newg}) and~(\ref{newgtilde}) with those numerical results.

The PT duration after $T$ drops below $T_n$ is quantified by $1/\beta$, where
\begin{equation}
	\beta \equiv \frac{1}{\Gamma_{\rm v}(t)}\frac{d\Gamma_{\rm v}(t)}{dt}\bigg|_{t=t_n},
\end{equation}
$\Gamma_{\rm v}$ is the decay rate per unit of three-dimensional volume of the false vacuum and $t_n$ is the value of the cosmic time $t$ when $T=T_n$. We will call $H_n$ the Hubble rate at $t=t_n$. 

\subsection{Sterile-neutrino dark matter  from reheating}\label{sec:Neutrini}

As an application of the results regarding reheating that we have presented in this section, we now discuss the production of sterile neutrinos from reheating. 

Several well-motivated extensions of the SM include sterile (right-handed) neutrinos $N_i$. One of these, $N_1$, can be sufficiently stable compatibly with the neutrino experimental data and can, therefore, be a good DM candidate~\cite{Canetti:2012kh} if an efficient production mechanism is available. 

This DM candidate can be produced via $\delta \chi$ decays  through a Yukawa coupling $y_1$, which enters the Lagrangian through the operator
\begin{equation}
    \Lag_{\rm DM}\equiv-\frac{1}{2}y_{1}\delta \chi N_1N_1+ \text{h.c.}.
\end{equation}
In this case the ratio between the DM energy density $\rho_{\rm DM}$ and the critical energy density $\rho_{cr}$ is  given by the branching ratio (BR)  of this $\delta\chi$ decay relative to the other decay channels:
\begin{equation}\label{abundance}
\Omega_{\rm DM}\equiv \frac{\rho_{\rm DM}}{\rho_{cr}}=\frac{s_0m_{N_1}}{3H_0^2/8\pi G_N}\text{BR}(\delta \chi \rightarrow N_1 N_1)\approx \frac{0.110}{h^2}\times\frac{m_{N_1}}{0.40\, \text{eV}}\text{BR}(\delta \chi\rightarrow N_1N_1),
\end{equation}
where $m_{N_1}$ is the mass of $N_1$, $s_0=g_{s0}T^3_02\pi^2/45$, $g_{s0}=43/11$
and $T_0$ and $H_0\equiv h\times 100$ km $\text{sec}^{-1}\text{Mpc}^{-1}$ are the present temperature and Hubble rate, respectively. A similar formula appears in the literature for the production of DM through inflaton decays (see e.g.~\cite{Cirelli:2024ssz} for a review).  Note that this is a non-thermal production mechanism. In general, for our setup~(\ref{abundance}) can be approximated by 
\begin{equation}
\label{eq:nonreducedDM}
    \Omega_{\rm DM}h^2\approx 0.110\times \frac{\chi_0}{0.40\, \text{eV}}\frac{\sqrt{\bar \beta \zeta_{N1}}}{2}\frac{\Gamma(\delta \chi\rightarrow N_1 N_1)}{\Gamma^{(2F)}+\Gamma^{(2S)}+\Gamma^{(2V)}+\Gamma^{(3S)}+\Gamma^{(SV)}}.
\end{equation}
where $\zeta_{N1}\equiv\frac{4m_{N1}^2}{m_\chi^2}=\frac{4y_1^2}{\bar \beta}$. When the relevant particles are only the SM ones and $N_1$ we can simplify the expression in~\eqref{eq:nonreducedDM}: since $\Gamma^{(2V)}$ and $\Gamma^{(SV)}$ dominate  the other decay rates in the denominators (see Fig.~\ref{fig:DRSMbeta-4}), 
\begin{equation}
\label{eq:nonreducedDMSM}
    \Omega_{\rm DM}h^2\approx 0.110\times \frac{\chi_0}{0.40\, \text{eV}}\frac{\sqrt{\bar \beta \zeta_{N1}}}{2}\frac{\Gamma(\delta \chi\rightarrow N_1 N_1)}{\Gamma^{(2V)}+\Gamma^{(SV)}}.
\end{equation}
When the three-body decay rate $\Gamma^{(SV)}$ is negligible and $\delta \chi\rightarrow N_1 N_1$ is kinematically open  we get 
\begin{equation}
\label{eq:reducedDMSM}
    \Omega_{\rm DM}h^2\approx0.110\times \frac{\chi_0}{0.40\,\text{eV}}\frac{\sqrt{\bar \beta 
 \zeta_{N1}}}{4}\frac{\zeta_{N1}(1-\zeta_{N1})^{3/2}}{3},
\end{equation}
where we used the small $\zeta_{VN}$ approximation to compute $\Gamma^{(2V)}$: the flat-direction field has to be significantly heavier than the SM fields for phenomenological reasons and this implies $\alpha\ll1$ and so, as we have seen,  the $\zeta$s in~(\ref{zeta1}) are small.
The factor of 3 in the denominator of~(\ref{eq:reducedDMSM}) is the number of massive vector bosons of the SM. 

The expression in~\eqref{eq:nonreducedDM} can be used in any RSB model featuring a sterile neutrino DM candidate. We will illustrate this in Sec.~\ref{Radiative electroweak and lepton symmetry breaking}, considering a well-motivated SM extension with RSB.

\section{A model of radiative EW and lepton symmetry breaking}\label{Radiative electroweak and lepton symmetry breaking}

As an application of the previous results we consider now an RSB SM extension that can account, unlike the SM, for neutrino oscillations, DM and baryon asymmetry. 

This model was constructed and studied  in~\cite{Salvio:2023ext} (see also Ref.~\cite{Iso:2009ss}) and its Lagrangian is given by
\begin{align}
    \label{eq:LBL}
  \Lag^{\rm ns}_{\rm matter}&=\Lag^{\rm ns}_{\rm SM}+D_\mu A^\dagger D^\mu A+\bar{N}_j i\slashed{D}N_j-\frac{1}{4}B'_{\mu\nu}B'^{\mu\nu}\\
    &+\left(Y_{ij}L_i\mathcal{H}N_j+\frac{1}{2}y_{ij}A N_iN_j+\text{h.c.}\right)-\lambda_a|A|^4 +\lambda_{ah}|A|^2|\mathcal{H}|^2, \nonumber
\end{align}
where the gauge group of the theory is the SM gauge group times a gauged $B-L$ factor, $U(1)_{B-L}$. Also,  $\Lag^{\rm ns}_{\rm SM}$  represents the classically scale-invariant part of the SM Lagrangian, $A$ is a complex scalar that is neutral under the SM gauge group, but with a non-vanishing value of $B-L$ required by the gauge invariance of the $AN_iN_i$ operator,  and the $L_i$ are the three families of SM lepton doublets. Similarly to what we have done for  $\mathcal{H}$, we can write
\begin{equation}
    A\equiv \frac{1}{\sqrt{2}}(a+i\eta_4),
\end{equation}
where $\eta_4$ is the would-be Goldstone boson of the broken gauge symmetry $U(1)_{B-L}$ when $A$ acquires a VEV (through RSB). Since we adopt the unitary gauge, $\eta_4$ only appears as the longitudinal degree of freedom of the extra massive gauge boson of the theory. 
In this model the covariant derivative can be written as
\begin{equation}
    D_\mu= \partial_\mu+ig_3T^\alpha G^\alpha_\mu+ig_2T^aW^a_\mu+ig_Y\mathcal{Y}B_\mu+i[g_m\mathcal{Y}+g'_1(B-L)]B'_\mu,
\end{equation}
which involve the gluons $G^\alpha_\mu$, the triplet of $W$ bosons $W^a_\mu$ as well as the gauge fields $B_\mu$ and $B_\mu'$ of $U(1)_Y$ and $U(1)_{B-L}$ (as usual $B'_{\mu\nu} \equiv \partial_\mu B'_\nu-\partial_\nu B'_\mu$) together with the respective generators $T^\alpha, T^a, {\mathcal Y}, B-L$ and gauge couplings $g_3, g_2, g_Y, g_1'$.
Here $g_m$ accounts for the mixing between $B_\mu$ and $B'_\mu$. 

The EWSB in this model can be analysed with the results of Sec.~\ref{sec:RSBEW} substituting $\lambda_\phi\to\lambda_a$ and $\lambda_{\phi a}\to \lambda_{ah}$.
In this model, however, a new massive gauge boson $Z'$ appears with mass $m_{Z'}=2|g'_1|\chi_0$. So  we assume that $\chi_0$ is large enough to safely avoid all the experimental constraints on $Z'$ bosons. So, to generate the EW scale $\lambda_{ah}$, and thus the mixing angle $\alpha$, has to be small (see Eq.~\eqref{eq:vev}). This means that $\chi$ is mostly $a$ and $H$ is mostly $h$.

Following~\cite{Asaka:2005pn,Canetti:2012kh} we take all right-handed neutrino Majorana masses (generated by the $AN_iN_i$ operator after $A$ gets a VEV) below the EW scale. This allows us to account not only for neutrino oscillations and baryon asymmetry but also for DM, as we will see. In this case, with good accuracy~\cite{Salvio:2023ext} 
\begin{equation}
\label{eq:betabl}
    \bar \beta=\frac{96 g_1'^4}{(4\pi)^2} 
    =\frac{8(4\pi)^2}{3\zeta_{Z'}^2}.
\end{equation}
In the second equality above we used the definition $\zeta_{Z'}\equiv 4 m_{Z'}^2/m_\chi^2$ or, equivalently, $\bar\beta= 16 g_1'^2/\zeta_{Z'}$ to eliminate $g_1'$.
Thus, the decay of $\delta \chi$ into two $Z'$ is kinematically forbidden in the perturbative regime. For example, for  $\bar \beta \sim 10^{-3}$, and assuming $\chi_0 \sim 10^{5}$ GeV, the dominant decay channel is $\delta \chi\rightarrow WW,ZZ$. For values of $\chi_0\gtrsim 10^{5}$ GeV also the channels $\delta \chi \rightarrow HWW, HZZ$ become relevant. Finally, \eqref{eq:reducedDMSM} becomes 
\begin{align}
      \Omega_{\rm DM}h^2 &\approx 0.110\times \frac{\chi_0}{0.40\,\text{eV}}2
    \pi\sqrt{\frac{2}{3}}\frac{1}{\zeta_{Z'}}\frac{\zeta_{N1}^{3/2}(1-\zeta_{N1})^{3/2}}{3}.
\end{align}
\begin{figure}[t!]
    \includegraphics[width=0.8\linewidth]{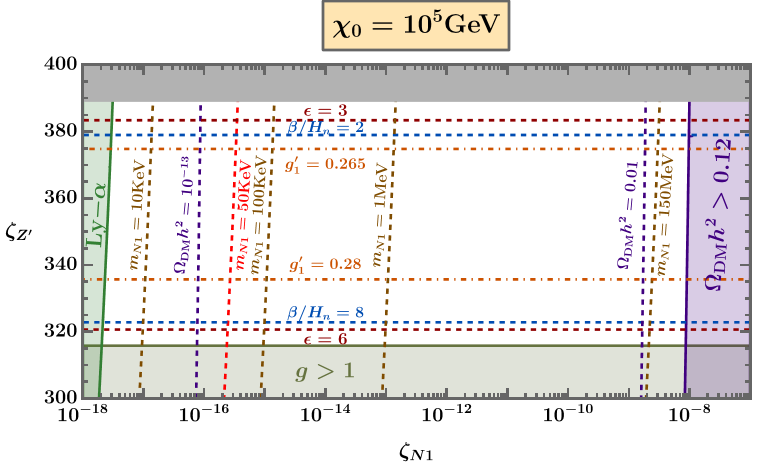}
    \caption{\it The parameter space for sterile-neutrino dark matter production via $\delta \chi$ decays in the model of Sec.~\ref{Radiative electroweak and lepton symmetry breaking}. $\beta/H_n$ has been computed with the method explained in Sec.~3.2.3 of~\cite{Salvio:2023ext}. We show in green the Lymann-$\alpha$ constraint on the mass of the non-thermal sterile neutrino DM, where we used the prescription presented in \cite{Cirelli:2024ssz}, in purple the overproduction constraint,  in dark gray the no-nucleation constraint and in dark green the region where the perturbation theory starts to be less accurate ($g>1$).}
  \label{fig:Neutrini5}
\end{figure}
In Fig.~\ref{fig:Neutrini5} we show the parameter space for  sterile-neutrino DM production. Also PBHs can be produced during PTs with model-independent mechanisms that work in the RSB scenario \cite{Liu:2021svg,Lewicki:2023ioy,Gouttenoire:2023naa,Banerjee:2024cwv}, and this can account for a fraction of DM. However, in the present model, using the results of~\cite{Banerjee:2024cwv}, we find that  the fraction of PBH DM produced is very small, less then $\sim 1\%$ for $\chi_0\sim 10^5$ GeV (see also Ref.~\cite{Baldes:2023rqv} for an independent analysis of PBH production in the same model and Ref.~\cite{Gouttenoire:2023pxh} for a related study). 

Interestingly, we find that a sterile neutrino produced through the decay of $\delta\chi$ can account for DM when its mass is around the $10^2$ MeV scale. This is significantly heavier than the sterile neutrino accounting for DM in the SM extended only through three right-handed neutrinos, where the production mechanisms are different (see e.g.~\cite{Canetti:2012kh}). In the latter model the sterile neutrino responsible for DM is not much heavier than 50 keV (we thus reported in red the line corresponding to $m_{N_1}=50$ keV in Fig.~\ref{fig:Neutrini5}). Note that  the Yukawa couplings $Y_{i1}$ can be taken to be sufficiently small to respect bounds from direct and indirect searches of this sterile neutrino.

We can check if the universe can be heated up to $T_{\rm eq}$ in this model using \eqref{eq:Trhrel2}, which is valid because, as we have seen,  the $\zeta$s in~(\ref{zeta1}) are small. For the relevant setup $\bar \beta \sim 10^{-3}$ and for the values of $\chi_0$ where the three-body decays are negligible with respect to the two-body decays into vectors (see Fig.~\ref{fig:DRSMbeta-4}),  the fast reheating condition \eqref{eq:Trhrel2} is always satisfied. Adding the contribution of the three-body decay needed at high $\chi_0$ can only weaken the constraint  on fast reheating. A previous analysis of reheating in this model was performed in Ref.~\cite{Baldes:2023rqv}, but without using the general results of Sec.~\ref{Reheating}. In our general analysis we considered all decay channels of $\delta\chi$, while  Ref.~\cite{Baldes:2023rqv} only considered the decays  $\delta\chi\to HH$ and $\delta\chi\to N_iN_j$. Another important difference is that DM here can be due to $N_1$, as an application of the general results of Sec.~\ref{sec:Neutrini}.

\section{Preheating}\label{Preheating}

There are cases when the reheating described in Sec.~\ref{Reheating} is not very efficient. This is typically the case when the RSB is a dark sector, meaning that it only features feeble interactions with the SM. When $\chi_0$ is well below the EW scale, $v$, we must be in this situation to satisfy the experimental bounds. We will now show how to reheat the universe in this situation. 

Two possible portal couplings come to mind: the Higgs portal in~(\ref{Hportal}) and a kinetic mixing between the $U(1)$ SM gauge boson and a dark photon (DP), a relatively light and somewhat hidden extra $U(1)$ gauge boson. 

The first portal does not generically lead to fast reheating in the above-mentioned situations: when $\chi_0\ll v$ such portal must be very small for phenomenological reasons, so that the energy density of $\delta\chi$ is typically transferred to other particles of the dark sector, which must be present and must have sizable couplings to $\delta\chi$ to trigger RSB. On the other hand, the $U(1)$ kinetic mixing, as we will see, can allow for an efficient transfer of the $\delta\chi$ energy density to the SM particles. The reason is essentially the fact that such gauge boson can  be the dark particle with the dominant coupling to $\delta\chi$. So the picture is that the $\delta\chi$ energy density is first transferred to the DP and then transferred to SM particles through the kinetic mixing.
 The former transfer cannot generically occur through the reheating mechanism described in Sec.~\ref{Reheating} because a sizable gauge coupling between the DP and $\delta\chi$ corresponds typically to a DP heavier than $\delta\chi$. We will see, however, that in this situation the preheating phase studied in~\cite{Dolgov:1989us,Traschen:1990sw,Kofman:1994rk,Kofman:1997yn} can do the job. In this section we will explore this possibility in the most general RSB setup. 

In the above-mentioned scenario the relevant part of the Lagrangian for the DP in the RSB sector is given by 
\begin{equation}
    \Lag_{\rm DP}=-\frac{1}{4}\mathcal{A}_{\mu\nu}\mathcal{A}^{\mu\nu}-\frac{\eta}{2}\mathcal{A}_{\mu\nu}\mathcal{F}^{\mu\nu}+\frac{1}{2}(D_\mu \phi)_a (D^\mu \phi)_a -V_{\rm ns}(\phi) ,
\end{equation}
where $\mathcal{A}_{\mu\nu}\equiv \partial_\mu \mathcal{A}_\nu-\partial_\nu\mathcal{A}_\mu$, the vector $\mathcal{A}_\mu$ is the gauge boson of the extra $U(1)$ symmetry, the second term is the kinetic mixing between the SM photon, $\mathcal{F}_\mu$, and $\mathcal{A}_\mu$,  $\mathcal{F}_{\mu\nu}\equiv \partial_\mu \mathcal{F}_\nu-\partial_\nu\mathcal{F}_\mu$ 
and $\eta$ is the mixing parameter.  

This mixing can be removed by performing the field redefinition $\{\mathcal{A}_\mu,\mathcal{F}_\mu\}\to\{A'_\mu,A_\mu\}$ given by
\be \left(
\begin{array}{c}
 \mathcal{A}_\mu  \\
\mathcal{F}_\mu  \\
\end{array}
\right)= \left(
\begin{array}{cc}
 \frac{1}{\sqrt{1-\eta ^2}} & 0 \\
 -\frac{\eta }{\sqrt{1-\eta ^2}} & 1 \\
\end{array}
\right)
\left(
\begin{array}{c}
 A'_\mu  \\
A_\mu  \\
\end{array}
\right). \label{redef}\ee
As a result, setting $\phi_a$ along the flat direction,
\be \frac{1}{2}(D_\mu \phi)_a (D^\mu \phi)|_{\phi_a = \nu_a \chi} \approx\frac12\partial_\mu \chi \partial^\mu\chi +\frac{e_d^2 \chi^2  A'_\mu A'^\mu}{2(1-\eta^2)}, \ee 
where $e_d$ is the gauge coupling between $\chi$ and the DP field and we have assumed that the DP is the particle with the dominant coupling to $\chi$ in the RSB sector. Inserting~(\ref{redef}) in the interaction between the SM photon and the electromagnetic current one finds an interaction between the DP and the electromagnetic current that is suppressed by $\eta$. So, $\eta$ must be small to satisfy the experimental constraints (see~\cite{Fabbrichesi:2020wbt,Graham:2021ggy} for reviews) and the DP background-dependent mass is $m_d(\chi)\approx |e_d| \chi$.

Therefore, the field equation of $A'_\mu$ always reads to good approximation
\be \frac1{\sqrt{|\det \hat g|}}\partial_\nu\left(\sqrt{|\det \hat g|}\, F'^{\mu\nu}\right)= m_d^2(\chi) A'^\mu, \label{ApEq}\ee
where $\det \hat g$ is the determinant of the spacetime metric, $F'_{\mu\nu}\equiv \partial_\mu A'_\nu-\partial_\nu A'_\mu$ and we have used the smallness of $\eta$. Note now that  $H_I^4$, where $H_I$ is the Hubble rate during the supercooling period, is always tiny compared to the potential density $\Delta V$ in any model of this sort: indeed, solving Einstein's equations leads to 
\be H_I = \frac{\sqrt{\bar\beta} \chi_0^2}{4\sqrt{3}\bp}, \label{HIf}\ee 
therefore, using~(\ref{DeltaV}), 
\be H_I^4 =\frac{\bar\beta}{144} \left(\frac{\chi_0}{\bp}\right)^4 \Delta V. \ee  
This result shows $H_I^4\ll \Delta V$ because $\bar\beta$ is loop suppressed and $\chi_0\ll\bp$ in order to remain in the regime of validity of the effective field theory of gravity. Therefore, after the false vacuum decay has taken place the energy stored in $\chi$ is not significantly dissipated by the expansion of the universe.

At some stage $\chi$ undergoes small oscillations around $\chi_0$.
When that happens, the dependence of $\chi$ on the cosmic time $t$ is given by 
\be \chi(t) =\chi_0+\Phi \sin(m_\chi t), \label{SmallOsc}\ee
 with $\Phi\ll\chi_0$, and the background dependent mass $ m_d(\chi)\sim |e_d|\chi_0$ is generically 
extremely large compared to $H_I$ (see Eq.~(\ref{HIf}) and recall that the DP is assumed to be the field with the dominant coupling to $\delta\chi$).  
We can, therefore, neglect the spacetime curvature in Eq.~(\ref{ApEq}) and then find the field equation
\be  \partial_\nu\, F'^{\mu\nu}= m_d^2(\chi) A'^\mu. \label{ApEOMosc}\ee
From the results obtained so far one can show that most of the energy stored in $\chi$ is transferred to $A'$ through preheating~\cite{Salvio:2023blb}. 

After this has happened, requiring $A'$ to be stable or sufficiently long lived as $\eta\to0$, the universe is reheated through the decays of $A'$ into SM fermions, which occur thanks to the interaction between $A'$ and the electromagnetic current. The decay rate into two leptons is of the form 
\begin{equation}
    \Gamma(A'\rightarrow\text{2 leptons})=\frac{(\eta e)^2m_d}{12\pi}\sqrt{1-\left(\frac{2m_l}{m_d}\right)^2}\left(1+\frac{2m_l^2}{m_d^2}\right)\Theta(m_d-2m_l),
\end{equation}
where $e$ is the electric charge, $m_l$ is the lepton mass and $m_d\equiv m_d(\chi_0)$. In the scenario we are interested in the decay rate of $A'$ into hadrons is also relevant: 
\begin{equation}
    \Gamma(A'\rightarrow \text{hadrons})=\frac{(\eta e)^2m_d}{12\pi}\sqrt{1-\left(\frac{2m_\mu}{m_d}\right)^2}\left(1+\frac{2m_\mu^2}{m_d^2}\right)R\left(\sqrt{s}=m_d\right),
\end{equation}
where $R\left(\sqrt{s}\right)\equiv\sigma(e^+e^-\rightarrow\text{hadrons},s)/\sigma(e^+e^-\rightarrow \mu^+\mu^-,s)$ \cite{ParticleDataGroup:2012pjm}.
One must require  $m_d\gtrsim 1$ MeV (twice the mass of the electron) in order for some of these decays to be kinematically allowed.

\begin{figure}[t!]
    \includegraphics[width=0.7\linewidth]{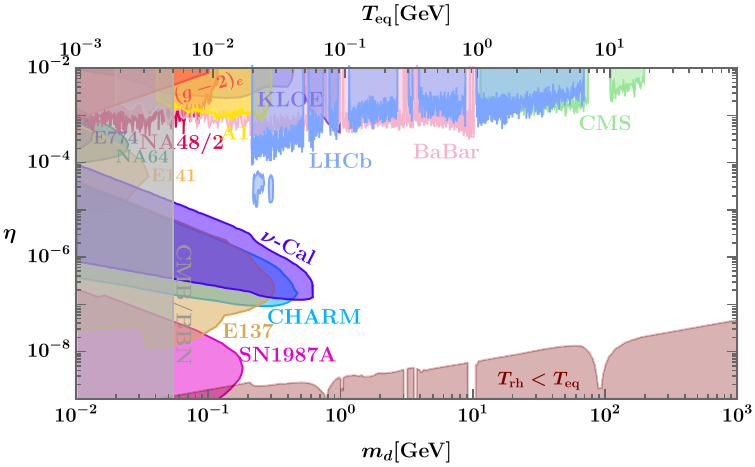}
    \caption{\it Region of the dark photon parameter space where fast reheating can occur. Here we set $g_*=100$ (only to report values of $T_{\rm eq}$, which, however, depend weakly on $g_*$) and  the relevant constraints on dark photons are included. Di-lepton searches with experiments at collider/fixed target: A1 \cite{Merkel:2014avp}, LHCb \cite{LHCb:2019vmc}, CMS \cite{CMS:2019kiy}, BaBar \cite{BaBar:2014zli}, KLOE \cite{KLOE-2:2011hhj,KLOE-2:2012lii,KLOE-2:2014qxg,KLOE-2:2016ydq}, NA48/2 \cite{NA482:2015wmo}, NA64 \cite{NA64:2018lsq}. Old beam dump: E774  \cite{Bross:1989mp}, E141 \cite{Riordan:1987aw}, E137 \cite{Bjorken:1988as, Batell:2014mga, Marsicano:2018krp}, $\nu$-Cal \cite{Blumlein:2011mv,Blumlein:2013cua}, CHARM \cite{Gninenko:2012eq}. Constraints from supernovae SN1987A \cite{Chang:2016ntp} and $(g-2)_e$ \cite{Pospelov:2008zw} are also included. Constraints on low reheating coming from BBN and CMB \cite{deSalas:2015glj,Hasegawa:2019jsa,Allahverdi:2020bys} are represented by the gray area, where we applied the constraint directly to the equivalence temperature, $T_{\rm eq}\gtrsim 5$ MeV. The region where $T_{\rm rh}< T_{\rm eq}$ is the region where the reheating is not fast. }\label{DPf}
    \end{figure}

In Fig.~\ref{DPf} we illustrate the region of the DP parameter space where fast reheating can occur together with the relevant
constraints on DPs.

\section{Conclusions}\label{Conclusions}

In this work we have shown how to reheat the universe after the supercooling period, which systematically occurs in the PTs associated with RSB. We have provided a general analysis, but we have also offered concrete examples when useful. 
\begin{itemize}
\item After introducing the general class of RSB theories, we have started by studying the reheating generated by the decays of $\delta\chi$ into SM particles and/or particles with sizable couplings to the SM. This reheating can be efficient when the RSB scale, $\chi_0$, is much larger than the EW scale, $v$. We have taken into account the leading interactions of $\delta\chi$ with all possible fields in a generic RSB theory and computed the corresponding decay rates (Secs.~\ref{Interactions} and~\ref{Decay Rates}).

To relate this calculation to the reheating temperature $T_{\rm rh}$, which must be the temperature of SM particles, in Sec.~\ref{sec:RSBEW} we have discussed the relation between EWSB and RSB when the SM is in the RSB sector: in that case also the EW scale has to be broken radiatively. We have illustrated this point in a prototypical setup when the Higgs field interacts with another scalar field through a quartic coupling. The Higgs field is a component of the flat-direction field $\chi$, which is responsible for RSB.

We have then used the above-mentioned results to provide explicit formul\ae~for $T_{\rm rh}$ and to give sufficient conditions for fast reheating in Sec.~\ref{Reheating}. In the same section we have also discussed when those conditions are compatible with the requirement that the bubbles of true vacuum are not diluted by the rapid expansion of the universe.

\item Furthermore, in Sec.~\ref{Reheating} we have illustrated how the decays of $\delta\chi$ can be a mechanism of DM production, considering the example of sterile-neutrino DM.

The above mentioned general results have then be applied to a concrete RSB SM extension with gauged $B-L$ and three right-handed (sterile) neutrinos. This model is particularly interesting because it is able to account for neutrino oscillations, baryon asymmetry and DM at the same time. We have shown that the lightest sterile neutrino can account for the entire DM abundance when its mass is around the $10^2$ MeV scale.  

\item When $\chi_0$ is not much larger than $v$, the mass and interactions of $\delta\chi$ are typically too small to reheat the universe only through decays of $\delta\chi$ into the SM sector and the RSB sector behaves as a dark sector. In this case, however, we have shown in Sec.~\ref{Preheating} that reheating always occur through preheating if a DP is the particle with the dominant coupling to $\chi$ in the RSB sector. In preheating the  production of particles interacting with $\chi$ occurs as a result of the time dependence of this field through parametric resonance and particles heavier than $\delta\chi$ can be produced. The full energy density stored in $\chi$ right after the PT is first transferred to the DP, which subsequently decays into SM fermions, reheating the universe. We have studied when this mechanism is sufficiently strong to guarantee fast reheating, $T_{\rm rh} = T_{\rm eq}$. 

\end{itemize}
\subsubsection*{Acknowledgments}
A.S.~thanks CERN for hospitality during the final stages of the project. This work has been partially supported by the Italian Ministry of University and Research (MUR) under the grant PNRR-M4C2-I1.1-PRIN 2022-PE2 Non-perturbative aspects of fundamental interactions, in the Standard Model and beyond F53D23001480006 funded by E.U. - NextGenerationEU.

\appendix

\section{Three-body phase space}\label{appendix:3ps}
In our case the amplitudes of the processes do not depend on the momentum, thus the phase space contribution and the dynamics are completely factorizable. 

Let us first consider  the $n$-body phase space in the center of mass frame of $\delta\chi$
\begin{equation}
\begin{aligned}
    \Omega_n(m_\chi,\{m_i\}) &\equiv \int \prod_{i=1}^n \frac{d^3p_i}{2E_i} 
    \delta\left(\sum_{i=1}^n \vec{p}_i\right)
    \delta\left(\sum_{i=1}^n E_i - m_\chi\right)= \\
    &= \int \frac{d^3p_n}{2E_n} 
    \int \prod_{i=1}^{n-1} \frac{d^3p_i}{2E_i} 
    \delta\left(\vec{p}_n+\sum_{i=1}^{n-1} \vec{p}_i\right)
    \delta\left(m_\chi - E_n-\sum_{i=1}^{n-1} E_i \right),
\end{aligned}
\end{equation}
where the $m_i$ are the masses of the final particles, the $E_i$ are the corresponding energies and the last integral is the phase space for $n-1$ particles with total momentum $-\vec{p}_n$ and total energy $m_\chi-E_n$, that we denote with $\Omega_{n-1}\left(,m_\chi-E_n, \{m_i\};-\vec{p}_n\right)$ , thus we can write
\begin{equation}
    \Omega_n\left(m_\chi,\{m_i\}\right)=\int \frac{d^3p_n}{2E_n}\Omega_{n-1}\left(m_\chi-E_n, \{m_i\};-\vec{p}_n\right).
\end{equation}
Since the phase-space integrals are Lorentz invariant we can rewrite $\Omega_{n-1}$ in the reference frame where the total momentum  and the total energy of the $n-1$ system is, respectively, zero and 
\begin{equation}
    \varepsilon =\sqrt{(m_\chi-E_n)^2-p_n^2}
\end{equation}
so that 
\begin{equation}
    \Omega_{n-1}(m_\chi-E_n, \{m_i\};-\vec{p}_n)=\Omega_{n-1}(\varepsilon, \{m_i\}),
\end{equation}
which implies the following recursion relation~\cite{Srivastava:1958ve}
\begin{equation}
    \Omega_n(m_\chi, \{m_i\})=\int \frac{d^3p_n}{2E_n}\Omega_{n-1}(\varepsilon, \{m_i\}).
\end{equation}
We can use this relation to reduce the three-body phase space calculation to a two-body one: 
\begin{equation}
    \Omega_3(m_\chi, m_i, m_j, m_k)=\int \frac{d^3p_k}{2E_k}\Omega_2(\varepsilon, m_i,m_j),\hspace{1cm} \varepsilon=\sqrt{(m_\chi-E_k)^2-p_k^2}.
\end{equation}
The expression for $\Omega_2(\varepsilon,m_i, m_j)$ is well known:
\begin{equation}
    \Omega_2(\varepsilon,m_i,m_j)=\frac{\pi \bar{p}_i}{\varepsilon},\hspace{1cm} \bar{p}_i=\frac{\sqrt{\varepsilon^2-(m_j-m_i)^2}\sqrt{\varepsilon^2-(m_j+m_i)^2}}{2\varepsilon}.
\end{equation}
As a result,
\bea
    &&\Omega_3(m_\chi,m_i, m_j, m_k) 
    = \int \frac{d^3p_k}{2E_k} \frac{\pi \bar{p}_i}{\varepsilon}= \nonumber  \\
    &&= \int \frac{4\pi^2 p_k^2 \, dp_k}{2E_k} 
    \frac{\sqrt{m_\chi^2 + m_k^2 - 2m_\chi E_k - (m_i - m_j)^2} \, 
          \sqrt{m_\chi^2 + m_k^2 - 2m_\chi E_k - (m_i + m_j)^2}}
         {2(m_\chi^2 + m_k^2 - 2m_\chi E_k)}. \nonumber 
\eea
The integration should be performed between a minimum and a maximum value of $p_k$, which we call $p_k^{\rm min}$ and $p_k^{\rm max}$, respectively. The minimum momentum  is $p_k^{\rm min}=0$, that occurs when particle $i$ and $j$ are emitted antiparallel with equal momenta. The maximum momentum is obtained when the other two particles are emitted parallel and with the same velocity, opposite to $\vec{p}_k$. This means that
\begin{equation}
    m_\chi=\sqrt{m_k^2+(p_k^{\rm max})^2}+\sqrt{(m_i+m_j)^2+(p_k^{\rm max})^2}.
\end{equation}
Solving for $p_k^{\rm max}$ we get 
\begin{equation}
    p_k^{\rm max}=\frac{\sqrt{m_\chi^2-(m_i+m_j-m_k)^2}\sqrt{m_\chi^2-(m_i+m_j+m_k)^2}}{2m_\chi}.
\end{equation}
Changing the integration variable, $dp_k=\frac{E_k}{p_k}dE_k$, and denoting with $E_k^{\rm max}$ the energy of the particle $k$ with momentum $p_k^{\rm max}$ we get 
 
\begin{align}
\Omega_{3}(m_\chi,m_i,m_j,m_k) &\equiv \int \frac{d^3p_i \, d^3p_j \, d^3p_k}{2E_i 2E_j 2E_k} \, \delta(\vec{p}_i + \vec{p}_j + \vec{p}_k) \, \delta(m_\chi - E_i - E_j - E_k) \nonumber \\
&=\pi^2 \int_{m_k}^{E_k^{\text{max}}} dE_k \,  \sqrt{E_k^2 - m_k^2} \nonumber \\
&\quad \times \frac{\sqrt{m_\chi^2 + m_k^2 - 2m_\chi E_k - (m_i - m_j)^2} \, \sqrt{m_\chi^2 + m_k^2 - 2m_\chi E_k - (m_i + m_j)^2}}{m_\chi^2 + m_k^2 - 2m_\chi E_k}. \nonumber
\end{align}

With the same procedure we can evaluate also the following phase space integral
\begin{align}
\label{eq:Omega34}
\Omega_{3}^{(4)}(m_\chi,m_j,m_k) &\equiv \int \frac{d^3k_i \, d^3k_j \, d^3p_k}{2E_i 2E_j 2E_k} \, (k_i\cdot k_j)^2\delta(\vec{k}_i + \vec{k}_j + \vec{p}_k) \, \delta(m_\chi - E_i - E_j - E_k) \nonumber \\
&= \frac{\pi^2}{4} \int_{m_k}^{E_k^{\rm max}} dE_k \, \sqrt{E_k^2 - m_k^2} \left(1 - \frac{2m_j^2}{m_\chi^2 + m_k^2 - 2m_\chi E_k}\right)^2 \nonumber \\
&\quad \times \sqrt{1 - \frac{4m_j^2}{m_\chi^2 + m_k^2 - 2m_\chi E_k}} \left(1 + \frac{m_k^2 - 2m_\chi E_k}{m_\chi^2}\right)^2 m_\chi^4,
\end{align}
where we assumed that two particles have equal mass ($m_i=m_j$). $\Omega_{3}^{(4)}$ appears in the computation of the decay into two vectors and one scalar. The two equal masses are the mass of the two identical vector bosons (see the second term in~(\ref{eq:LVint})).

The dimensionful  phase space integrals can be written as
\begin{align}\label{eq:ODim}
    \Omega_{3}(m_\chi,m_i,m_j,m_k)&\equiv\omega_{3}(\bar \beta,\alpha_i,\alpha_j,\alpha_k)\chi_0^2, \nonumber\\
    \Omega_{3}^{(4)}(m_\chi,m_j,m_k)&\equiv\omega_{3}^{(4)}(\bar \beta, \alpha_j,\alpha_k)\chi_0^6,
\end{align}
where   $m_i\equiv\alpha_i\chi_0$ and the dimensionless part of all the phase-space integrals of interest is given by
\bea
  \hspace{-1cm}  \omega_{3}(\bar \beta, \alpha_i,\alpha_j,\alpha_k)&\equiv& \pi^2\int^{x^{\rm max}_k}_{\alpha_k}dx_k\,\sqrt{x_k^2-\alpha_k^2} \sqrt{1-\frac{(\alpha_i-\alpha_j)^2}{\bar \beta+\alpha_k^2-2\sqrt{\bar \beta}\,x_k}}\sqrt{1-\frac{(\alpha_i+\alpha_j)^2}{\bar \beta+\alpha^2_k-2\sqrt{\bar \beta}\, x_k}},\nonumber \\
    \omega_{3}^{(4)}(\bar \beta,\alpha_j,\alpha_k)&\equiv&\frac{\pi^2}{4}\bar \beta^2\int^{x_k^{\rm max}}_{\alpha_k} dx_k\sqrt{x_k^2-\alpha_k^2}\left(1-\frac{2\alpha_j^2}{\bar \beta+\alpha_k^2-2\sqrt{\bar \beta}x_k}\right)^2\sqrt{1-\frac{4\alpha_j^2}{\bar \beta + \alpha_k^2-2\sqrt{\bar \beta}x_k}}\nonumber \\
 &&\times \left(1+\frac{\alpha_k^2-2\sqrt{\bar \beta}x_k}{\bar \beta}\right)^2,\nonumber
\eea
where $x^{\rm max}_k$ is defined by
\begin{equation}
   E_k^{\rm max}=\sqrt{(p_k^{\rm max})^2+m_k^2}\equiv x^{\rm max}_k\chi_0,  \hspace{0.5cm}  p_k^{\rm max}=\frac{\sqrt{\bar \beta}}{2}\sqrt{1-\frac{(\alpha_i+\alpha_j-\alpha_k)^2}{\bar\beta}}\sqrt{1-\frac{(\alpha_i+\alpha_j+\alpha_k)^2}{\bar \beta}}\chi_0. \nonumber
\end{equation}

 \vspace{1cm}
\footnotesize
\begin{multicols}{2}

\end{multicols}

\end{document}